\documentclass[aps,pra,amsfonts,amssymb,epsfig,twocolumn,superscriptaddress]{revtex4}

\usepackage{graphics,graphicx}
\usepackage{amsmath,bbm}
\usepackage{amstext}
\usepackage{amssymb}
\usepackage{color}

\def\beq{\begin{equation}}
\def\eeq{\end{equation}}
\def\bea{\begin{eqnarray}}
\def\eea{\end{eqnarray}}
\providecommand{\openone}{\leavevmode\hbox{\small1\kern-3.8pt\normalsize1}}

\begin{document}

\title{Homogeneous MERA states: an information theoretical analysis}

\author{V. Giovannetti}
\affiliation {NEST CNR-INFM \& Scuola Normale Superiore, Piazza
dei Cavalieri 7, I-56126 Pisa, Italy}
\author{S. Montangero}
\affiliation {Institut f\"ur Quanteninformationsverarbeitung,
  Universit\"at Ulm, D-89069 Ulm, Germany}
\author{M. Rizzi}
\affiliation{Max-Planck-Institut f\"{u}r Quantenoptik, Hans-Kopfermann-Str. 1, D-85748 Garching, Germany}
\author{Rosario Fazio}
\affiliation {NEST CNR-INFM \& Scuola Normale Superiore, Piazza
dei Cavalieri 7, I-56126 Pisa, Italy}

\date{\today}

\begin{abstract}

Homogeneous Multi-scale Entanglement Renormalization Ansazt (MERA) state have been recently introduced 
to describe quantum critical systems. 
Here we present an extensive analysis of the properties of
such states by clarifying the definition of their transfer super-operator whose structure is studied  within a
informational theoretical approach.
Explicit expressions for computing the expectation values of symmetric observables are given both in the case
of finite size systems and in the thermodynamic limit of infinitely many particles.  
\end{abstract}

\pacs{03.67.-a,05.30.-d,89.70.-a}

\maketitle

\section{Introduction}
The physics of strongly interacting many-body quantum system is 
central in many areas of physics. 
 Our ability of simulating  them is based on the possibility 
to find an efficient description of their ground state.
This is the case, for example, of White's Density Matrix 
Renormalization Group~\cite{DMRG1} which can be recasted in terms of Matrix Product States
(MPS)~\cite{FNW,OR,VIDAL1,VPC,rev}.   
Such representations  are characterized by a 
tensor  decomposition of the many-body wave-function which allows one  
to efficiently compute all the  relevant observables of the system (e.g. energy, 
local observables,  and correlation functions), and  to reduce the 
effective number of parameters over which the numerical optimization  needs to 
be performed. MPS fulfill these requirements and  can be used to describe faithfully
the ground states of not critical, short range one-dimensional many-body Hamiltonians. 
However MPS are not efficient  in providing an accurate
description in other relevant situations, i.e. when the system is
critical, in higher physical dimensions or if the model possesses
long-range couplings. Several proposals have been put forward to
overcome this problem.  
Projected Entangled Pair States~\cite{peps} generalize MPS in dimensions 
higher than one.  Weighted graph states~\cite{wgs} can deal with long-range correlations. 
Here we focus on a solution recently proposed by 
Vidal~\cite{mera} who introduced a tensor structure based on the so called
Multiscale Entanglement Renormalization Ansatz (MERA). The MERA tensor network
satisfies both the above efficiency requirements  and accommodates 
the scale invariance typical of critical systems~\cite{meraalg,meraapp}.
The relevance of this approach might 
represent a major breakthrough in our simulation capabilities~\cite{2DMERA} 
and motivates an intensive study of the MERA -- e.g. see Refs.~\cite{vidallong,DEO,NOSTRO,nostro1,ALTER}. 
In Ref.~\cite{NOSTRO} some of us described a  connection between the MERA 
and the theory of completely positive quantum maps~\cite{BENGZY}.
In the context of  {\em homogenous} MERA's 
(i.e. MERA's formed by identical layers of 
tensors -- see below for details) 
this permits to introduce a transfer matrix formalism
 in the same spirit as it has been done for MPS~\cite{VPC,OR,WOVC}, while providing 
new tools to compute physical observables using MERA's. 
As a result  a connection between the critical exponents governing the decay of 
two-points correlation functions and the eigenvalues of the MERA transfer matrix was identified yielding 
a simple method for determining the properties of critical many-body systems in the thermodynamic 
limit~\cite{ALTER,nostro1}.

In the present paper
we shall review
some of the results introduced in Ref.~\cite{NOSTRO} providing explicit
derivations and clarifying the underlying mathematical aspects of the problem.  
In particular we formalize an important property of homogeneous MERA's by 
presenting two theorems that allow one to evaluate the expectation values of symmetric observables 
(including translationally invariant Hamiltonians) in terms of a unique MERA transfer super-operator.
Furthermore the thermodynamic limit of the MERA states is analyzed clarifying the condition under which
such limit exists. 

The paper is organized as follows.  Sec.~\ref{sec1} is  devoted to review the basics of the MERA tensor network. We   discuss their causal cone structure and introduce the subset of homogenous MERA states.
Even though most of this material can be found elsewhere~\cite{mera,meralun} we decided to 
insert it here to make the paper self-consistent. This Section introduces also a new theoretical tool   (the {\em causal shadows} of the MERA) which will play a fundamental role in the subsequent derivation. 
Sec.~\ref{sec:Qu}  is the central core of the paper: here we  analyze the quantum channel description of MERA's showing 
 how global quantities such us energy, average
magnetization, etc. of a homogeneous MERA state  can be described in terms of a single super-operator
(the average QuMERA channel). Also the MERA transfer operator is defined by moving in the 
Liouville representation~\cite{ROYER,BENGZY} 
(the latter is reviewed in Appendix~\ref{Liou}).
In Sec.~\ref{sec:term} and Sec.~\ref{wwwsecscali} we then discuss the 
thermodynamic limit of a MERA state and the scaling behavior of its two-point 
correlation functions by using general properties of {\em mixing} 
quantum channels~\cite{NJP,RAG,GOHM,TDV} and exploiting the 
spectral properties of the associated QuMERA channel.
The paper ends with the conclusions in Sec.~\ref{sec:conc}.

\begin{figure}[h!]
\begin{center}
\includegraphics[scale=.65]{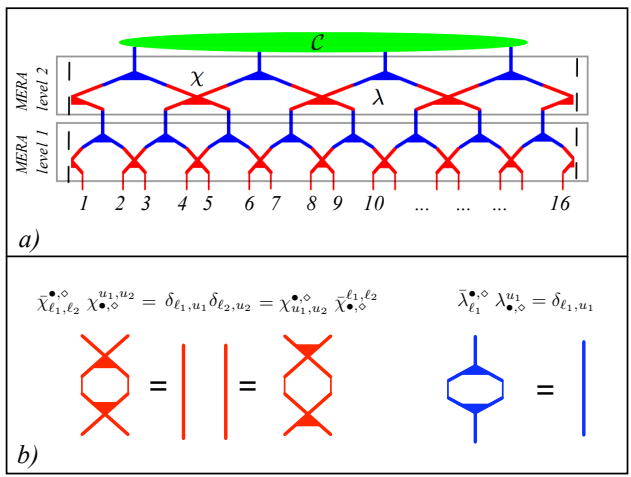}
\caption{{\em a)} Graphical representation of a typical one dimensional MERA tensor network~\cite{mera} for a  many-body system of $N=16$ sites. 
The red elements correspond to the disentaglers tensors $\chi$, the blue elements are the isometry tensors $\lambda$,
and the green element ${\cal C}$ is the {hat} of the MERA.
Any two joined legs from any two distinct nodes indicate saturation of the associated indices of the corresponding
tensor~\cite{mera}. The dashed line indicate periodic boundary conditions 
(i.e. the right-most $\chi$ re-emerge on the left of the graph). 
 Alternative MERA decompositions can be obtained by 
reordering the links of the graph, e.g. see Ref.~\cite{ALTER}.
{\em b)} Representation of the contraction rules of Eqs.~(\ref{condi1}), (\ref{condi2})  imposed on $\chi$ and $\lambda$ (here the 
inverted elements  represents their adjoints counterparts).}
 \label{fig1}
\end{center}
\end{figure}

\section{The MERA network}\label{sec1}
In this section we give a brief, self-consistent review of 
the basics of the MERA tensor decomposition which was introduced by Vidal in Ref.~\cite{mera}. 
The only new element is the formalization of  the notion 
the Causal Shadows presented in Sec.~\ref{sub:cs}.

\subsection{Basics} 
Consider a many-body quantum system $S$ composed by
$N=2^n$ sites of dimension $d$ (qudits). Its pure states 
can be expressed as
\begin{eqnarray}
|\Psi\rangle = \sum 
 \; {\cal T}_{\ell_1,\ell_2,\cdots, \ell_N} 
 | \xi_{\ell_1} ,\xi_{\ell_2} , \cdots ,  \xi_{\ell_N}\rangle\;, \label{neweq1}
 \end{eqnarray}
 where for $j\in\{1,\cdots,N\}$ and $\ell \in\{ 1, \cdots, d\}$ the vectors
 $|\xi_{\ell}\rangle_j\in{\cal H}_d$ form the computational basis of the $j$-th 
 system site and where the type-$
 \mbox{\tiny{$\left(\begin{array}{c} 0 \\ N
 \end{array}\right)$}}$ tensor 
 ${\cal T}_{\ell_1,\ell_2,\cdots, \ell_N}:= 
 \langle  \xi_{\ell_1},\cdots, \xi_{\ell_N} |  \Psi\rangle$ are the associated 
 probability amplitudes.
The MERA representation~\cite{mera} assumes a 
decomposition of ${\cal T}$ in terms of a collection  of smaller, finite size tensors, which
differently from the  linear MPS structure~\cite{FNW}, are 
 organized in  a complex two-dimensional graph. An explicit example is shown 
in  Fig.~\ref{fig1}.
Here the links emerging from the lowest part of the graph represent the $N$  {\em physical indices}
of ${\cal T}$ associated with  the sites of $S$.
The nodes of the graph instead
 represent  tensors. They are divided in three groups: the 
 type-$
 \mbox{\tiny{$\left(\begin{array}{c} 2 \\2
 \end{array}\right)$}}$ {\em disentangler} tensors $\chi$    of
 elements ${\chi}^{u_1,u_2}_{\ell_1,\ell_2}$ 
represented by the red
 Xs;  
 the type-$
 \mbox{\tiny{$\left(\begin{array}{c} 1 \\2
 \end{array}\right)$}}$
 tensors ${\lambda}$  of elements $\lambda^{u_1}_{\ell_1,\ell_2}$ 
represented by the blue inverted Ys;
 and  the type-$
 \mbox{\tiny{$\left(\begin{array}{c} 0 \\ 4
 \end{array}\right)$}}$
tensor $\cal C$ of elements
${\cal C}_{\ell_1,\ell_2,\ell_3,\ell_4}$, represented by the green blob.
As shown in Fig.~\ref{fig1} the  $\chi$'s, the $\lambda$'s are coupled in
 together to form a triangular structure 
with  ${\cal C}$ as the closing element of the top: any two joined legs from any two distinct 
nodes indicate saturation of the associated indices. 
Consequently, apart from the hat,  ${\cal T}$ 
is  written as a
network of $O(N)$ smaller tensors
organized in $m=\log_2(N) -2$
different layers which we enumerate from the bottom of the graph.

For a generic MERA  the  tensors entering the decomposition may differ from node to node and their
indexes may have arbitrary (finite) dimensions. Under these conditions any state of $S$ can be 
represented as in Fig.~\ref{fig1} by a proper choise of the $\chi$'s and the $\lambda$'s.
 In the following however we will restrict the analysis to the special class 
of {homogeneous} MERA states in which all the $\chi$'s and the $\lambda$'s
entering the decomposition 
are identical and in which all the indexes of the graph have the same dimension $d$~\cite{NOTAnew1}. 
With this choice the MERA 
identifies a much narrower but more treatable subset of many-body quantum 
states. The interest  in such subset is motivated by the fact that 
homogeneous MERA's possess an intrinsic scale invariance symmetry built in
which is typical of critical, translationally invariant systems.
In particular by removing the first $m'$ layers from a $N$-site homogenous MERA state $|\Psi\rangle$ 
we obtain smaller versions of such vector constructed with only $N/2^{m'}$ sites which, however, 
in the limit of sufficiently large  $N$ (thermodynamic limit) still preserve the same correlations of the original one.
Such a symmetry is believed~\cite{mera,NOSTRO,nostro1,ALTER} to be sufficient for characterizing (at least approximatively) the ground state properties of critical, translationally invariant Hamiltonians,
making homogenous MERA states optimal candidates for their numerical simulations.

\subsection{Causal cones} \label{subsec:caus}
What really makes the MERA decomposition a convenient one is the 
assumption that 
the tensors composing the graph satisfy special 
contraction rules~\cite{mera} -- see Fig.~\ref{fig1} part b). Specifically one requires the following identities
\begin{eqnarray} \label{condi1}
[ \bar{\chi} \cdot \chi ]_{\ell_1,\ell_2}^{u_1,u_2} 
 = [ \chi \cdot \bar{\chi}]_{\ell_1,\ell_2}^{u_1,u_2}   &=& \delta_{\ell_1}^{u_1}\delta_{\ell_2}^{u_2} \;,
  \\
\big[ \bar{\lambda} 
\cdot \lambda\big]_{\ell}^{u} 
&=& \delta_{\ell}^{u}\;,\label{condi2}
\end{eqnarray}
where $\delta$ is the Kronecker delta, ``$\cdot$" represents upper-lower  contraction of consecutive
tensors~\cite{NOTACONTRACT},    and where
 $\bar{\chi}$ and $\bar{\lambda}$ are the adjoints of $\chi$ and $\lambda$
 defined by
\begin{eqnarray}
\bar{\chi}^{u_1,u_2}_{ \ell_1, \ell_2} =({\chi}^{
  \ell_1, \ell_2}_{u_1,u_2})^*\;,
\qquad
 \bar{\lambda}^{u_1,u_2}_{ \ell_1} =({\lambda}^{ \ell_1 }_{ u_1,
  u_2})^*\;.
  \end{eqnarray}
Expressed in operator language, Eqs.~(\ref{condi1}), (\ref{condi2}) 
imply that $\chi$ and $\lambda$ can be interpreted, respectively,  as unitary transformation acting on two qudit sites
and as an isometry that maps one qudit into two qudits.

\begin{figure}[t!]
\begin{center}
\includegraphics[scale=.7]{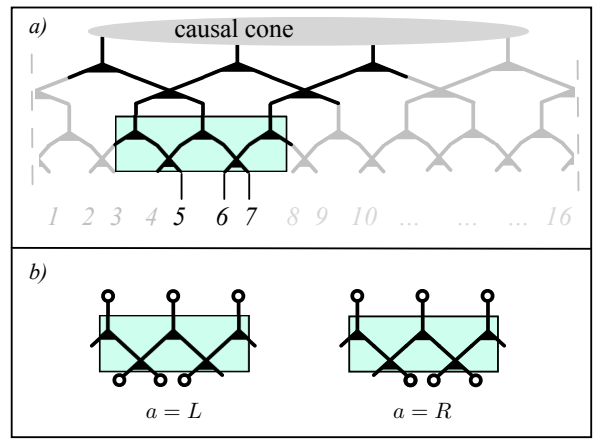}
\caption{{\em a)} The black elements of the graph represent the tensor ${\cal Q}_k$   of  the 
causal cone associated with the triple formed by the 
$5$-th, $6$-th and $7$-th sites of the system -- see  Eq.~(\ref{QTENSOR}). It can be identify by connecting the triple with the MERA's hat via percolation starting from the bottom of the graph. Thanks to the contraction rules~(\ref{condi1}) and (\ref{condi2})
the remaining $\chi$'s and $\lambda$'s (in gray) do not contribute when evaluating the expectation values of
observables which act locally on the triple. The light box underlines the tensor ${\cal M}$  of Eq.~(\ref{M5}). Here $N=16$. 
{\em b)} The two alternative ways in which a tensor ${\cal M}$ can enter in ${\cal Q}_k$: the empty circle represent
the links that connect with the neighboring elements of the cone. 
} \label{figura2}
\end{center}
\end{figure}

Under these constraints 
 each triple formed by three consecutive sites of the system 
is associated with a {\em causal cone} (CC) identified via percolation -- see Fig.~\ref{figura2}:
 Only the $\chi$'s and the $\lambda$'s belonging to  the CC  can contribute not trivially 
 in the evaluation of the of expectation values of the local  observables acting on such triple~\cite{NOTA3}. 
 This is an important property  of the MERA which allows one to  reduce the number of contractions
 that need to be performed when evaluating expectation values on $|\Psi\rangle$ from $O(N\log_2 N)$ to only $O(\log_2 N)$, exponentially simplify 
 the complexity of the calculation~\cite{mera}. 
 As a result given $\hat{{A}}_k$ an observable acting not trivially on the triple of sites $k-1,k$ and $k+1$, we can
express its expectation value as
\begin{eqnarray}\label{expvalue}
\langle \Psi |\hat{{A}}_k|\Psi\rangle =
(\bar{\cal C}_{k} \cdot \bar{\cal Q}_k^{(m)}) 
\cdot {\cal A}_k \cdot ({\cal Q}_k^{(m)}  \cdot {\cal C}_{k})\;,
 \end{eqnarray}
 where ${\cal A}_k$ is the tensor associated with the operator  $\hat{A}_k$~\cite{NOTATENSO},  
 ${\cal Q}_k^{(m)}$ is  the 
 tensor associated with the CC of the triple $k$, and ${\cal C}_k$ is the hat tensor with $k$ specifying which of its $4$ lower indexes couple to ${\cal Q}_k^{(m)}$
 (as before ``$\cdot$" and the ``$\bar{...}$" represent upper-lower index contraction and the adjoint operation).
 ${\cal Q}_k^{(m)}$ is obtained by properly cascading  $m$ copies of the following  type-$
 \mbox{\tiny{$\left(\begin{array}{c} 3 \\ 6
 \end{array}\right)$}}$ tensor ${\cal M}$,  
 \begin{eqnarray}
 [{\cal M} ]_{\ell_1,\cdots,\ell_6}^{u_1, u_2, u_3} : = \lambda_{\ell_1,\circ}^{u_1} \;  \chi_{\ell_2,\ell_3}^{\circ,\bullet} \; \lambda_{\bullet,\diamond}^{u_2} \;   \chi^{\diamond, \star}_{\ell_4,\ell_5}\;  \lambda_{\star,\ell_6}^{u_3}\;,
 \label{M5}
 \end{eqnarray}
 where, as in Ref.~\cite{NOSTRO}, for easy of the notation we  use
 typographic symbols $\circ, \bullet, \diamond, \star$ to 
indicate summation over the corresponding index. The way such tensors couple with each other and with ${\cal A}_k$
and ${\cal C}_k$ is specified by their position within the cone and ultimately depends upon the 
location on the triple in $S$. 
As shown in the figure there are two possibilities. Specifically the ${\cal M}$ tensor of the $(m'-1)$-th  layer of the CC
can saturate its upper indexes $u_1,u_2,u_3$ either with  the lower indexes $\ell_2,\ell_3,\ell_4$ (modality $L$) or with 
the lower indexes $\ell_3,\ell_4,\ell_5$ (modality $R$) of the ${\cal M}$  tensor of the $m'$-th layer. 
We distinguish the two cases by assigning a label  $a\in\{ L,R\}$ 
to each element of ${\cal Q}_k^{(m)}$  -- see Fig.~\ref{figura2}{\em b)}. 
With this choice we can now write
\begin{eqnarray}
 {\cal Q}_k^{(m)} = 
 {\cal M}_{a_{1}^{(k)}} \cdot   {\cal M}_{a_{2}^{(k)}} \cdot \ldots \cdot  {\cal M}_{a_{m}^{(k)}} \label{QTENSOR}\;,
 \end{eqnarray}
where, for $m'\in\{1,\cdots, m\}$ and $k\in \{ 1,\cdots, N\}$ the index $a_{m'}^{(k)}\in\{ L,R\}$
specifies which lower indexes ${\cal M}_{a_{m'}^{(k)}}$ uses to connect with  ${\cal M}_{a_{m'-1}^{(k)}}$ 
(or with ${\cal C}_k$  and ${\cal A}_k$  if $m'=1$ or $m'=m$).

\begin{figure}[t!]
\begin{center}
\includegraphics[scale=.78]{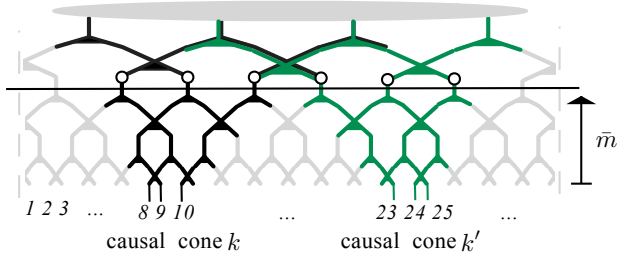}
\caption{The black {\em and} green elements represent the causal cone structure associated to 
the product $\hat{A}_k \otimes \hat{B}_{k'}$
(here $N=32$, $k=9$, and $k'=24$): it is formed by merging the CC of the $k$ triple (black elements)
with the CC of the $k'$ triple (green elements) which intercept at the $(\bar{m}+1)$-th MERA level.
The empty circles  describe the $6$-qudits quantum state on which the two single-triple CC's operate upon.
It is associated with the tensor ${\cal X}_{kk'}$ of Eq.~(\ref{expvalue1}) and with the density matrix $\hat{\sigma}_{{k,k'}}$
of Eq.~(\ref{corrimp}).
} \label{figura3}
\end{center}
\end{figure}

Analogous  simplifications occur also for non-local observables. Of particular interest are the
2-point correlation functions of the form
 $\langle \Psi| \hat{A}_k \otimes \hat{B}_{k'} |\Psi \rangle$ with $\hat{A}_k$ and $\hat{B}_{k'}$ being (local) operators which act not trivially on the triples formed by the sites $k-1$, $k$, $k+1$ and $k'-1$, $k'$, $k'+1$, respectively.
In this case the contraction rules~(\ref{condi1}), (\ref{condi2}) determine  a joint CC for the sites $k$, $k'$
formed by 
two single-triple CC (one for each triple),  
 which intercept at the $\bar{m}+1$  MERA layer (counting from the bottom of the graph) with~\cite{NOSTRO},
 \begin{eqnarray}
 \bar{m} = \mbox{int}[\log_2 |k-k'|]-1 \;, \label{barm}
 \end{eqnarray}
 see Fig.~\ref{figura3}.
 This allows us to express $\langle \Psi |\hat{A}_k\otimes \hat{B}_{k'} |\Psi\rangle$ as
 \begin{eqnarray}\label{expvalue1}
\left(\bar{\cal X}_{kk'} \cdot (\bar{\cal Q}_k^{(\bar{m})} \bar{\cal Q}_{k'}^{(\bar{m})} ) \right)
\cdot ({\cal A}_k {\cal B}_{k'} ) \cdot  \left(
({\cal Q}_k^{(\bar{m})} {\cal Q}_{k'}^{(\bar{m})} ) \cdot {\cal X}_{kk'} \right),
 \end{eqnarray}
where ${\cal A}_k {\cal B}_{k'}$ is the tensor associated with $\hat{A}_k\otimes \hat{B}_{k'}$, 
${\cal Q}_k^{(\bar{m})} {\cal Q}_{k'}^{(\bar{m})}$ is tensor which describes the CC up to the $\bar{m}$-th layer (it is given by the product of two independent single triple CC~(\ref{QTENSOR})),   
and where ${\cal X}_{kk'}$ describes the convolution of the MERA hat with the remaining part of the CC (i.e. the part above the $\bar{m}$-th layer).

\begin{figure}[t!]
\begin{center}
\includegraphics[scale=.75]{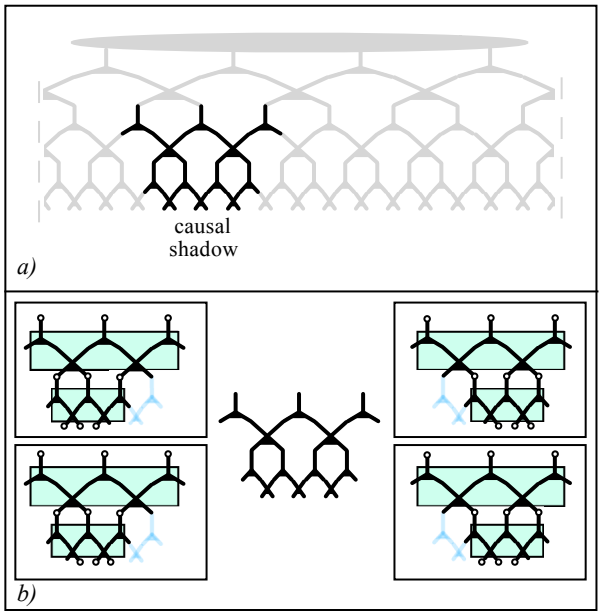}
\caption{{\em a)} The black links  represents an example of first neighboring causal shadow
of depth 2 associated with three consecutive links of the
second layer of a $N=32$-MERA. {\em b)}Tensor of  a causal shadow of depth 2 associated with a triple
 The insets show its decomposition in terms of products~(\ref{QTENSOR}): starting from the top-left corner and moving clockwise we have ${\cal M}_L\cdot {\cal M}_L$,
${\cal M}_R\cdot {\cal M}_R$, ${\cal M}_L\cdot {\cal M}_R$ and ${\cal M}_R\cdot {\cal M}_L$.
} \label{figurashadow}
\end{center}
\end{figure}
\subsection{Causal  Shadows} \label{sub:cs}

A notion which is complementary to CC is provided by  what we define the 
{\em Causal Shadows} (CS's) of the MERA.
If  CC's play a fundamental role in the calculation of expectation values on the MERA state,
the CS's  are fundamental in simplifying the analysis of symmetric quantities as will be clear in the next section. 

Given a certain set of links ${\cal L}$ of the $m'$-th MERA layers, we define its associated CS  
 as the set of all CC's that 
 allow one to reach elements of ${\cal L}$ (and only those) 
 from the physical indexes of the MERA (i.e. the bottom of the graph) and define $m'$ its depth --  
more precisely CS is the set of $\chi$'s and  $\lambda$'s belonging to such CC's. 
A trivial example of CS is obtained by considering $m'=m$ (upmost MERA layer) and identifying ${\cal L}$ with the set
of 4 emerging links: in this case the CS includes all $\chi$'s and $\lambda$'s of the MERA.  
Less trivial examples are shown in Figure~\ref{figurashadow}. 
For each CS we can clearly associate a tensor (this is the tensor formed by the $\chi$'s and the $\lambda$'s of CS)
and a set of physical indexes (this is the set  of physical indexes that are attached to the CS tensor).
Of particular interest for us are the CS's associated with triples of links as those shown in Fig.~\ref{figurashadow}.
A simple combinatorial analysis shows that the physical indexes of such causal shadows  
 contain $2^{m'}+2$ contiguous elements with $m'$ being the depth of the CS. Most importantly one can verify each one of such CS tensor contain all possible $2^{m'}$ sequences  formed by
combining $m'$  tensors ${\cal M}$ either with modality $L$ or with modality $R$: this is a trivial consequence of the fact
that the CS includes all possible paths~(\ref{QTENSOR}) which ends in the selected triple --- see Fig.~\ref{figurashadow} b) for an example.

\section{QuMERA channels}\label{sec:Qu}
 
 A better insight and more 
compact expressions for the expectation values  on homogeneous MERA's
can be obtained by moving in super-operator language~\cite{NOSTRO,ALTER,nostro1}.  Within this approach 
the tensors entering  the causal cones associated with a specific subset of MERA's sites are re-organized to
form concatenations of certain quantum channels (the {\em QuMERA channels}) whose definitions and properties depend explicitly upon the selected subset of sites. 
In the following we will review this approach and provide an explicit proof of two
Theorems that allow one to derive a simple analytical  expression  
for the average quantities computed  on the state $|\Psi\rangle$ associated with the MERA.

\subsection{Local observables}
Let us start considering the case of observables 
$\hat{A}_k$ operating on the triple $k$ formed by the neighboring sites
$k-1,k$ and $k+1$ for which Eq.~(\ref{expvalue}) applies (single-sites observables are trivially included 
as a special case).
We notice that, depending on the value of $a$, the ${\cal M}_a$ tensors  contributing to~(\ref{QTENSOR})
can be associated with  two families of operators 
$\{\hat{L}_{r}\}_{r}$ and $\{ \hat{R}_{r}\}_{r}$ 
acting on the
Hilbert space ${\cal H}_d^{\otimes 3}$ and labeled through the composed index
${r} :=( r_1, r_2,r_3)$ with $r_{1,2,3}$ being $d$-dimensional. In the computational
basis they are 
defined by the matrices 
\begin{eqnarray}
\langle \xi_{u_1} , \xi_{u_2},  \xi_{u_3} |\hat{L}_{r} |
 \xi_{\ell_1} , \xi_{\ell_2} , \xi_{\ell_3}\rangle =[{\cal M}]^{u_1,u_2,u_3}_{r_1,
  \ell_1,\ell_2,\ell_3,r_2,r_3},  \\
 \langle \xi_{u_1} , \xi_{u_2},  \xi_{u_3} |\hat{R}_{r} |
 \xi_{\ell_1} , \xi_{\ell_2} , \xi_{\ell_3}\rangle
=[{\cal M}]^{u_1,u_2,u_3}_{
 r_1, r_2,\ell_1,\ell_2,\ell_3,r_3 },
 \end{eqnarray}
and are related through a reshuffling $\Pi$ of the
input and output qudits, i.e. 
\begin{eqnarray}
\hat{R}_{r}= \Pi( \hat{L}_{r}) :=
\hat{P} \hat{L}_{r} \hat{P}^\dag \label{sim}\;,
\end{eqnarray}
where $\hat{P} = \hat{P}^\dag$ is the unitary transformation which exchanges 
the first and the third qudit.
Most importantly, according to the contraction rules~(\ref{condi1}), (\ref{condi2}), the
sets $\{\hat{L}_{r}\}_{r}$ and $\{ \hat{R}_{r}\}_{r}$ 
satisfy the following normalization conditions 
\begin{eqnarray}
\sum_{r} \hat{L}_{r}\hat{L}_{r}^\dag = \hat{\openone}^{\otimes 3} = \sum_{r} 
\hat{R}_{r} \hat{R}_{r}^\dag\;,
\end{eqnarray}
with $\hat{\openone}$ being the identity operator of ${\cal H}_d$.
Therefore $\{ \hat{L}_{r}\}_{r}$  can be used to identify
a completely positive, unital, not necessarily trace preserving super-operator
$\Phi_H^{(L)}$~\cite{BENGZY},
which transforms the linear operators $\hat{A}$ of ${H}_d^{\otimes 3}$ according to the following expression
\begin{eqnarray} \label{defPHIL}
\Phi_H^{(L)}(\hat{A}) = \sum_{r} \hat{L}_{r} \hat{A} \hat{L}_{r}^\dag\;.
\end{eqnarray}
Analogously $\{ \hat{R}_{r}\}_{r}$   defines the map $\Phi_H^{(R)}$ 
which is related with $\Phi_H^{(L)}$ through the identity
\begin{eqnarray}
\Phi_H^{(R)} = \Pi \circ \Phi_H^{(L)} \circ \Pi\;,
\end{eqnarray}
where  ``$\circ$" indicates the  composition of super-operators~\cite{NOTAIMPO}. We will refer to such operators
(and to their compositions) as QuMERA  channels (i.e.  quantum MERA channels).
We also introduce the  following vector of 
 ${\cal H}_d^{\otimes 4}$,
\begin{eqnarray}
|\Psi_{\text{hat}}\rangle :=  \sum_{\ell_1,\ell_2,\ell_3,\ell_4}  
{\cal C}_{\ell_1,\ell_2,\ell_3,\ell_4}  \; |\xi_{\ell_1},
 \xi_{\ell_2},\xi_{\ell_3},\xi_{\ell_4}\rangle\;,
 \end{eqnarray}  
which  can be assumed to be  normalized
(thanks to Eqs.~(\ref{condi1}), (\ref{condi2}) the norm of $|\Psi_{\text{hat}}\rangle$ 
and $|\Psi\rangle$ coincides). For $j\in\{1,2,3,4\}$ we then  introduce
  $\hat{\rho}_C^{(j)}: = \mbox{Tr}_{j} [  |\Psi_{\text{hat}}\rangle\langle \Psi_{\text{hat}}|]$ the
 reduced density matrices  obtained by tracing $|\Psi_{\text{hat}}\rangle\langle \Psi_{\text{hat}}|$ over one of its $j$-th qudits, e.g.
\begin{eqnarray}
\hat{\rho}_C^{(1)} &: =& 
\sum_{\ell_1} \sum_{\ell_2,\ell_3,\ell_4}
\sum_{\ell'_2,\ell'_3,\ell'_4}
{\cal C}_{\ell_1,\ell_2,\ell_3,\ell_4} 
 {\cal C}^*_{\ell_1,\ell'_2,\ell'_3,\ell'_4}\nonumber  \\&&\times
|\xi_{\ell_2},\xi_{\ell_3},\xi_{\ell_4}\rangle \langle \xi_{\ell'_2},\xi_{\ell'_3},\xi_{\ell'_4}|\;,
\end{eqnarray}
(assuming $|\Psi_{\text{hat}}\rangle$ to be symmetric under permutations the $\hat{\rho}_C^{(j)}$ becomes identical).
With these definitions one can  finally cast the expectation value~(\ref{expvalue}) as  
\begin{eqnarray}\label{explicit}
\langle \Psi|  \hat{A}_k |\Psi\rangle = \mbox{Tr} [ \hat{\rho}_C^{(j_k)} \; {\Phi_H^{(a_{m}^{(k)})} \circ \cdots  \circ  \Phi_H^{(a_{1}^{(k)})}} (\hat{A})],
\end{eqnarray}
where  $j_k\in \{ 1,2,3,4\}$ 
and where (enumerating from the lower MERA level of Fig.~\ref{fig1})  $\Phi_H^{(a)}$ is either the map $\Phi_H^{(L)}$ or $\Phi_H^{(R)}$ associated with
the corresponding element ${\cal M}_a$ of the causal cone~(\ref{QTENSOR}). Notice also
that in the rhs we have removed the label $k$ from $\hat{A}_k$ since there is no longer the need to specify
over which triple the operator is acting (the trace runs in fact over 3 qudits only).
The explicit value of $j_k$ as well as the sequence of maps  entering Eq.~(\ref{explicit}) depend upon 
$N$ and  $k$: for instance in the case shown in Fig.~\ref{figura2} (i.e. $N=16$, $k=6$) the super-operator
 sequence is $\Phi_H^{(L)}\circ  \Phi_H^{(R)}$ while $j=4$.
Exploiting Hilbert-Smidth duality Eq.~(\ref{explicit}) can  also be 
written as
 \begin{eqnarray}\label{imp}
 \langle \Psi|\hat{A}_k  |\Psi \rangle
=  \mbox{Tr} [ {\Phi^{(a_{1}^{(k)})} \circ \cdots  \circ  \Phi^{(a_{m}^{(k)})}}  (\hat{\rho}_C^{(j_k)}) \;   \hat{A}]  \;,
\end{eqnarray}
with  $\Phi^{(a)}$  being the super-operator $\Phi_H^{(a)}$
in Schr\"{o}dinger picture (i.e. its adjoint with respect to the Hilbert-Smidth product). This is a completely positive, trace preserving (CPT) channel~\cite{BENGZY} 
whose operator sum representation is provided by the operators $\{ \hat{L}_r^\dag\}_r$ (if $a=L$)
or by the operators $\{ \hat{R}_r^\dag\}_r$ (if $a=R$).
Since Eq.~(\ref{imp}) holds for all observables $\hat{A}_k$ we can finally conclude that 
\begin{eqnarray}
{\Phi^{(a_{1}^{(k)})} \circ \cdots  \circ  \Phi^{(a_{m}^{(k)})}}  (\hat{\rho}_C^{(j_k)}) = \hat{\rho}_k\;, \label{matricerid}
\end{eqnarray}
with $\hat{\rho}_k$ being the reduced density matrix of $|\Psi\rangle$ associated with the triple $k$.

Building up from these results we now
present a  theorem which formalize previous observations~\cite{NOSTRO,nostro1,ALTER,meralun}: 
 \newline
 
 {\bf Theorem 1:} {\em Let $\hat{A}^{(s)}$ be the symmetric version of the local operator $\hat{A}$, i.e.
 $\hat{A}^{(s)} : = \frac{1}{N} \sum_{k=1}^N \hat{A}_k$. Its expectation value on the homogeneous 
 MERA state $|\Psi\rangle$ can be computed as
  \begin{eqnarray}\label{imp1111}
\langle \Psi| \hat{A}^{(s)} |\Psi \rangle = 
   \mbox{{\em Tr}} [ {\Phi^m}  (\hat{\rho}_C) \;   \hat{A}]  \;,
\end{eqnarray}
where 
$\Phi^m := \Phi \circ\Phi \circ \cdots \circ \Phi$
 with $\Phi$ being the equally weighted mixture of $\Phi^{(R,L)}$, 
\begin{eqnarray} \label{QUMERA1}
\Phi:= \frac{\Phi^{(R)} + \Phi^{(L)}}{2} \;,
\end{eqnarray}
 and $\hat{\rho}_C:= \sum_{j=1}^4 \hat{\rho}^{(j)}_C/4$.}
\newline 

{\em Proof:} From the definition of $\hat{{A}}^{(s)}$ and from Eq.~(\ref{matricerid}) we can write 
\begin{eqnarray}
\label{proof1}
&&\langle \Psi| \hat{{A}}^{(s)} |\Psi \rangle = \frac{1}{N} \sum_{k=1}^N 
   \mbox{{Tr}} [ \hat{\rho}_k   \;  \hat{{A}}_k]  \nonumber \\
   &&=  \frac{1}{N} \sum_{k=1}^N 
   \mbox{{Tr}} [{\Phi^{(a_{1}^{(k)})} \circ \cdots  \circ  \Phi^{(a_{m}^{(k)})}}  (\hat{\rho}_C^{(j_k)})  \;  \hat{A}]
   \;.
\end{eqnarray}
We now notice that when varying $k$, 
${\Phi^{(a_{1}^{(k)})} \circ \cdots  \circ  \Phi^{(a_{m}^{(k)})}}$ spans all possible $m$-long sequences of  $\Phi^{(L)}$ 
and $\Phi^{(R)}$. As a matter fact each of such sequence is counted four times 
(one for each possible values of $j_k$). 
Remembering that $m= \log_2 N-2$
we can thus reorder the summation over $k$  on the last term of Eq.~(\ref{proof1}) 
as follows 
 \begin{eqnarray}
\label{proof2}  \tfrac{1}{4 \times 2^m}
\sum_{j=1}^4\sum_{\vec{a}\in \{ L,R\}^m}  
   \mbox{{Tr}} [{\Phi^{(a_{1})} \circ \cdots  \circ  \Phi^{(a_{m})}}  (\hat{\rho}_C^{(j)})  \;  \hat{{A}}] \;,
\end{eqnarray}
where $\vec{a}$ is the string $(a_1,a_2,\cdots, a_m)$. 
Using then the identity
\begin{eqnarray} 
\label{identityimpo}\frac{1}{2^m} \!
\sum_{\vec{a}\in \{ L,R\}^m} 
\Phi^{(a_{1})} \circ \cdots  \circ  \Phi^{(a_{m})}   =
 \left(\frac{\Phi^{(L)} + \Phi^{(R)}}{2}\right)^m,
 \end{eqnarray}
 this finally gives Eq.~(\ref{imp1111}). $\blacksquare$
   \newline
   
{\em Remark:--} An alternative proof can be constructed by expressing the involved tensor contraction in terms
of the CS's associated with all possible triples of links of the MERA hat and by exploiting the fact that each of such CS's
contains all possible combination of ${\cal M}_L$ and ${\cal M}_R$ (i.e. of $\Phi^{(L)}$ and $\Phi^{(R)}$).
\newline

An important application of this theorem is obtained by considering
the expectation value of translationally invariant Hamiltonians $\hat{H}$
with first nearest- and second nearest-neighbors coupling, i.e. 
\begin{eqnarray}
\hat{H}: = \sum_{i=1}^N \left( \hat{H}_{i-1,i,i+1}^{(3)} + \hat{H}_{i, i+1}^{(2)} + \hat{H}_i^{(1)} \right)\;, \label{HMI}
\end{eqnarray}
with  $\hat{H}_{i-1,i, i+1}^{(3)}$, $\hat{H}_{i,i+1}^{(2)}$ 
describing 3-body and 2-body interactions, and 
with $\hat{H}_i$ being local terms (here $i=0$ and $i=N+1$ are identified with $i=N$ and $i=1$ respectively to enforce the proper periodic conditions).
This can be expressed as 
\begin{eqnarray}
\hat{H}= \sum_{k=1}^N \hat{h}_{k} \label{HMI1}\;,
\end{eqnarray}
where 
\begin{eqnarray}
\hat{h}_k: &=& \hat{H}_{k-1,k,k+1}^{(3)}+
\frac{\hat{H}^{(2)}_{k-1,k} + \hat{H}^{(2)}_{k,k+1} }{2} \nonumber \\
&&+ \frac{\hat{H}^{(1)}_{k-1} + \hat{H}^{(1)}_{k}+\hat{H}^{(1)}_{k+1}}{3} \;,
\end{eqnarray}
 is the Hamiltonian terms associated with the triple 
formed by the sites $k-1,k$, and $k+1$. We can thus interpret $\hat{H}/N$ as the symmetric
version of the local observable $\hat{h}$. Therefore from 
Eq.~(\ref{imp1111}) follows  the identity
 \begin{eqnarray}\label{imp2222}
\frac{\langle \Psi| \hat{H} |\Psi \rangle}{N} = 
   \mbox{Tr} [ {\Phi^m}  (\hat{\rho}_C) \;   \hat{h}]  \;.
\end{eqnarray}
This expression shows that, for homogeneous MERA's,  the evaluation of the   average 
energy per  site  $E={\langle \Psi| \hat{H} |\Psi \rangle}/{N}$
of a generic translational invariant Hamiltonian $\hat{H}$ can be expressed in terms of the
channel $\Phi$ and of the (symmetric) reduced density operator of the MERA's hat.

\subsection{Two-points correlation functions}\label{sec:corr}

Let now focus on the correlation functions of the form 
$\langle \Psi| \hat{A}_k \otimes \hat{B}_{k'} |\Psi \rangle$ with 
$\hat{A}_k$ and $\hat{B}_{k'}$ being generic observable operating on the $k\neq k'$ triples
respectively.
Applying the derivation of the previous section  to Eq.~(\ref{expvalue1}) we can write  
 \begin{eqnarray}\label{corrimp}
&&\langle \Psi| \hat{A}_k \otimes \hat{B}_{k'} |\Psi \rangle  \\ \nonumber
&&= \mbox{Tr} [ \Phi^{(b_{1}^{(k,k')})} \circ \cdots  \circ  \Phi^{(b_{\bar{m}}^{(k,k')})} (\hat{\sigma}_{kk'}) \;  ( \hat{A} \otimes \hat{B})]  \;,
\end{eqnarray}
where for $m'\in \{1,\cdots, \bar{m}\}$ the
$\Phi^{(b_{m'}^{(k,k')})}$ are $6$-qudits QuMERA channels (Schr\"{o}dinger picture)
 associated with the first $\bar{m}$ levels of the casual cone of the sites $k$, $k'$ (see Fig.~\ref{figura3}), while
$\hat{\sigma}_{kk'}$ is a $6$-site density matrix obtained by ``evolving" the MERA hat with the remaining part of the
causal cone and tracing out some of the links (which one is indicated by the indexes ${k,k'}$ and ultimately depends upon the interception between the two independent causal cones of $k$ and $k'$ 
-- see Fig.~\ref{figura3} for an example: here the $6$-sites of $\hat{\sigma}_{kk'}$ are indicated by  empty circles).
An explicit expression for $\Phi^{(b_{m'}^{(k,k')})}$ is obtained as follows
\begin{eqnarray} \label{2sites}
\Phi^{(b_{m'}^{(k,k')})} : = \Phi^{(a_{m'}^{(k)})} \otimes \Phi^{(a_{m'}^{(k')})}
 \end{eqnarray}
with $\Phi^{(a_{m'}^{(k)})}$ and $\Phi^{(a_{m'}^{(k)})}$ being the single triple maps associated with the causal cones of
$k$ and $k'$ respectively. Thus, depending on $k$, $k'$ the map $\Phi^{(b_{m'}^{(k,k')})}$ will be one of the following four channels,
$\Phi^{(L)} \otimes \Phi^{(L)}$, $\Phi^{(L)} \otimes \Phi^{(R)}$, $\Phi^{(R)} \otimes \Phi^{(L)}$
 or $\Phi^{(R)} \otimes \Phi^{(R)}$.
As in the case of Eqs.~(\ref{imp}), (\ref{matricerid}) we can then use the fact the Eq.~(\ref{corrimp})
holds for all possible two-triple observables to conclude that the joint state of the triples $k,k'$ can be expressed as
\begin{eqnarray}
\Phi^{(b_{1}^{(k,k')})} \circ \cdots  \circ  \Phi^{(b_{\bar{m}}^{(k,k')})} (\hat{\sigma}_{kk'}) =
\hat{\rho}_{k,k'} \;, \label{jointstate}
\end{eqnarray}
with $\hat{\rho}_{k,k'}$ the reduced density matrix of the MERA state $|\Psi\rangle$ associated with such triples.

The evaluation of Eq.~(\ref{corrimp}) is in general quite complicated as it requires  to compose four  different maps 
in a specific order determined by the involved CC's  (of course for some clever choice of $k$ and $k'$ such a sequence could be relatively simple to compute).  
One would be tempted to solve this problem by ``symmetrizing" the two-point operator as in the local observable case
(e.g. replacing $\hat{A}_k \otimes \hat{B}_{k'}$ with $ \frac{1}{N} \sum_{k=1}^N 
 \hat{A}_k \otimes \hat{B}_{k + \Delta k}$, where $\Delta k= k'-k$). Unfortunately this is not sufficient, the reason being
 ultimately related with the fact that even for homogenous MERA the state $|\Psi\rangle$ is in general NOT translational invariant for finite $N$~\cite{NOTAPERO}.
 
\begin{figure}[t!]
\begin{center}
\includegraphics[scale=.78]{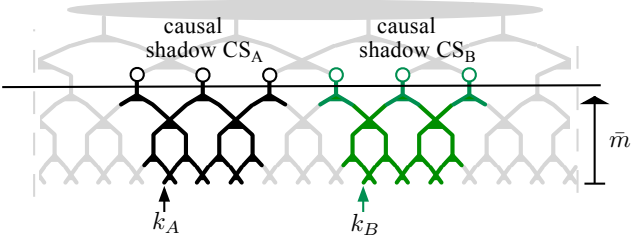}
\caption{The black and green elements represent the causal shadows CS$_A$ and CS$_B$ respectively.
The arrows indicate their leftmost sites $k_A$ and $k_B$
(here $N=32$, $k_A=8$, and $k_B=20$). The empty circles describe the 6-sites described by the density matrix $\hat{\sigma}^{(k_A,\Delta_k)}$ of Eq.~(\ref{corrimp1111}).
} \label{figureshd2}
\end{center}
\end{figure}

 One way to circumvent this is to exploit the CS structure to enforce a ``local" symmetrization of $\hat{A}_k \otimes \hat{B}_{k'}$. Specifically, let  $\hat{A}$  being a generic observable acting on a triple of qudits. Consider then a causal shadow 
 CS$_A$ of depth $\bar{m}$
 characterized by  $\bar{M}= 2^{\bar{m}} +2$
 physical indexes $\{\ell_{k_A},\ell_{k_A+1},\cdots, \ell_{k_A+\bar{M}}\}$ which will be grouped in a sequence
 of consecutive $\bar{M}-2$ triples
 labelled as $k_A+1, k_A+2, \cdots, k_A+{\bar{M}}-1$ (here $k_A$ is the leftmost physical index
 of CS$_A$ whose explicit value depends upon the position of the CS$_A$ within the MERA --- see Sec.~\ref{sub:cs} and Fig.~\ref{figureshd2}).
  We define  the {\em shadow} operator as the average of $\hat{A}$ over the triples of CS$_A$, i.e. 
  \begin{eqnarray}
\hat{A}_{k_A}^{(s)} : = \frac{1}{2^{\bar{m}}}\sum_{k=k_A+1}^{k_A+1 +2^{\bar{m}}} \hat{A}_{k}\label{shadowops}\;,
 \end{eqnarray}
where $\hat{A}_{k}$ is the operator $\hat{A}$ acting on the $k$-th triple of CS$_A$.
Consider then a second operator $\hat{B}$ and a second CS$_B$  of depth $\bar{m}$ which is first neighbor with CS$_A$
(that is the rightmost index at the top  of CS$_A$ is first neighbor with the leftmost index
at the top of CS$_B$, or vice-versa -- see Fig.~\ref{figurashadow}). It is worth noticing that the physical sites of CS$_A$
and CS$_B$ are separated by  a distance which is  exponentially large in $\bar{m}$. Indeed such distance can be easily computed as 
\begin{eqnarray}
\Delta_k := k_B-k_A =2( 2^{\bar{m}} -1) .\label{Delta}
\end{eqnarray}
with $k_B$ being the leftmost physical triple of CS$_B$ (here for the sake of simplicity we assume $k_B>k_A$).
Another important property is the  fact that given CC$_A$ and CC$_B$ 
causal cones associated with CS$_A$ and CS$_B$, respectively, they  are independent up to the level $\bar{m}$ but 
{\em intercept} at the level $\bar{m}+1$. As a matter of fact one can easily verify that  {\em all} couple of CC's that share this property enters in CS$_A$ and CS$_B$.

Also for CS$_B$  we define now
a shadow operator
\begin{eqnarray}
\hat{B}_{k_B}^{(s)} : = \frac{1}{2^{\bar{m}}}\sum_{k'=k_B+1}^{k_B+1+2^{\bar{m}}} \hat{B}_{k'}\label{shadowops1}\;,
 \end{eqnarray}
and consider the joint shadow observable 
\begin{eqnarray} \label{jointope}
\hat{{AB}}^{(s)}_{k_A,\Delta_k}:= \hat{A}_{k_A}^{(s)} \otimes \hat{B}_{k_B}^{(s)}\;.
\end{eqnarray}
This is a symmetrized version of $\hat{A}_{k} \otimes \hat{B}_{k'}$ obtained by averaging locally over all
possible choices of $k$ and $k'$ whose causal cone intercept in the same points of the MERA. 
\newline
 
 {\bf Theorem 2:} {\em Let $\hat{{AB}}^{(s)}_{\Delta_k}$ be defined as in Eq.~{\em (\ref{jointope})}.
 Its expectation value on the homogeneous 
 MERA state $|\Psi\rangle$ can be computed as
  \begin{eqnarray}\label{corrimp1111}
\langle \Psi| \hat{{AB}}^{(s)}_{k_A,\Delta_k} |\Psi \rangle = 
   \mbox{{\em Tr}} [ ({\Phi}^{\bar{m}} \otimes \Phi^{\bar{m}})  (\hat{\sigma}^{(k_A,\Delta_k)})    (\hat{A} \otimes \hat{B} )
   ] 
\end{eqnarray}
where $\bar{m}$ is the depth of the causal shadows CS$_A$ and CS$_B$,
$\Phi$ as in Eq.~{\em (\ref{QUMERA1})},  and where now $\hat{\sigma}^{(k_A,\Delta_k)}$ is the $6$-qudit density operator
associated with the upper indexes of the two causal shadows -- see Fig.~\ref{figureshd2}.}
\newline 

{\em Proof:} The proof proceeds as in the case of Theorem 1. First of all we use Eq.~(\ref{jointope}) to 
write 
 \begin{eqnarray}\label{corproof}
\langle \Psi| \hat{{AB}}^{(s)}_{k_A,\Delta_k} |\Psi \rangle = \left(\frac{1}{2^{\bar{m}}}\right)^2 
\sum_{k,k'} 
   \mbox{{Tr}} [ \hat{\rho}_{kk'}\;   (\hat{A}_k \otimes \hat{B}_{k'} )
   ]  \;,
\end{eqnarray}
where the summation is performed over the physical sites of CS$_A$ and CS$_B$ and where  $\hat{\rho}_{k,k'}$ is the joint  reduced density matrix of $|\Psi\rangle$ associated with the triples $k$ and $k'$. The latter can be expressed as in 
Eq.~(\ref{jointstate}) with $\Phi^{(b_{m'}^{(k,k')})}$ being the 2-site QuMERA channel~(\ref{2sites}) 
associated with the causal cones of the triples $k$ and $k'$. Most importantly 
in our case the $6$-qudits density matrix $\hat{\sigma}_{k,k'}$ is {\em independent} from $k$ and $k'$:
in fact it represents the state of the $(\bar{m}+1)$-th level that is attached with the CC's associated with the triples $k$
and $k'$ and by construction all the CC's belonging to a given CS intercept in the same points. Exploiting this 
we identify $\hat{\sigma}_{k,k'}$ with $\hat{\sigma}^{(k_A,\Delta_k)}$ of Eq.~(\ref{corrimp1111}).
Furthermore the (independent) average over  $k$ and $k'$ allows one to
 generate all possible $\bar{m}$-long  sequences of $\Phi^{(L)}$, $\Phi^{(R)}$ in both CS's. 
We can thus write the rhs of Eq.~(\ref{corproof}) as
\begin{eqnarray}\label{corproof1}
 &&\sum_{k,k'}  \mbox{{Tr}} [ \frac{
 [\Phi^{(a_{1}^{(k)})} \otimes \Phi^{(a_{1}^{(k')})} ] \circ \ldots \circ
 [\Phi^{(a_{\bar{m}}^{(k)})} \otimes \Phi^{(a_{\bar{m}}^{(k')})} ]}{(2^{\bar{m}})^2}
\nonumber \\
&& \qquad \qquad \qquad \qquad \times \nonumber (\hat{\sigma}^{(k_A,\Delta_k)})
     (\hat{A} \otimes \hat{B}) ]  \\
&&   =  \mbox{{Tr}} 
   [\sum_{k} 
 \tfrac{  [\Phi^{(a_{1}^{(k)})} \circ  \ldots \circ \Phi^{(a_{\bar{m}}^{(k)})}]} {2^{\bar{m}}}
  \otimes 
  \sum_{k'} \tfrac{
  [\Phi^{(a_{1}^{(k')})}  \circ \ldots \circ \Phi^{(a_{\bar{m}}^{(k')})} ]}
  {2^{\bar{m}}} 
  \nonumber  \\ && \qquad \qquad \qquad \qquad \times \nonumber (\hat{\sigma}^{(k_A,\Delta_k)})
     (\hat{A} \otimes \hat{B} ) ]\;.
  \end{eqnarray}
 Equation~(\ref{corrimp}) finally follows by the identity~(\ref{identityimpo}).
$\blacksquare$
   \newline

{\em Remark:--} We can further symmetrize the correlation function by averaging Eq.~(\ref{jointope}) with respect to their
absolute position within the MERA (keeping the relative distance among the CS's constant), i.e.  by replacing  
$\hat{{AB}}^{(s)}_{k_A,\Delta_k}$ with 
\begin{eqnarray}
\hat{{AB}}^{(s)}_{\Delta_k}:=\frac{1}{N}
\sum_{k_A=1}^N \hat{{AB}}^{(s)}_{k_A,\Delta_k}\;, \label{ABSIM}
\end{eqnarray}
where $k_A$ is the leftmost triple of CS$_A$.  
With this choice Eq.~(\ref{corrimp1111}) still applies by substituting    $\hat{\sigma}^{(k_A,\Delta_k)}$ with its
average counterpart 
$\hat{\sigma}:= \sum_{k_A=1}^{N} \hat{\sigma}^{(k_A,\Delta_k)} /N$, i.e. 
  \begin{eqnarray}\label{corrimp2}
\langle \Psi| \hat{{AB}}^{(s)}_{\Delta_k} |\Psi \rangle = 
   \mbox{{Tr}} [ ({\Phi}^{\bar{m}} \otimes \Phi^{\bar{m}})  (\hat{\sigma}) \;   (\hat{A} \otimes \hat{B} )
   ]  \;.
\end{eqnarray}

In Sec.~\ref{wwwsecscali} we shall see how Eq.~(\ref{corrimp2}) 
can be used to determine the scaling behavior of two-point correlations function of a MERA.

\subsection{The MERA transfer operator} \label{sec:to}

Theorem 1 and 2 formally show  that in  extracting  local or correlated  (average) quantities of $|\Psi\rangle$ one can focus on a single CPT map $\Phi$  obtained by averaging over all possible triple-sites QuMERA channels.
 This leads us to identification of a {\em transfer operator}  for the MERA~\cite{NOSTRO} in close similarity to what happens in the case of MPS (e.g. see Ref.~\cite{WOVC}). 

The idea  is to 
move   to the  Liouville space representation~\cite{BENGZY,ROYER} in which 
by ``doubling" the degree of freedom of the system,  the super-operators
are represented by matrices, and the operators by vectors (see Appendix~\ref{Liou} for details).
Specifically this is done by fixing an orthonormal  basis $\{ |i\rangle; i\}$~\cite{NOTAELEMENTO} on 
 the three qubits space ${\cal H}_d^{\otimes 3}$ and  associating 
 to each operator $\hat{A}$ of such system 
 a vector of $| \hat{A} \rangle\rangle \in {\cal H}_d^{\otimes 3} \otimes {\cal H}_d^{\otimes 3}$ 
 defined by 
\begin{eqnarray}
| \hat{A} \rangle\rangle := \sum_{ii'}  \langle i| \hat{A}|i'\rangle \;\;  |i\rangle\otimes | i' \rangle 
\label{dueee}.
\end{eqnarray}
According to this formalism the QuMERA channel ${\Phi}$ of Eq.~(\ref{QUMERA1}) 
can now be  described by the
 {\em transfer operator}~\cite{NOSTRO} acting on
${\cal H}_d^{\otimes 3} \otimes {\cal H}_d^{\otimes 3}$ defined by
\begin{eqnarray}\label{transfoperator}
\hat{E}_\Phi := \frac{1}{2} \sum_{r} \left[ \hat{L}_r^\dag \otimes  \hat{L}_r^T + 
\hat{R}_r^\dag \otimes  \hat{R}_r^T \right] \;, 
\end{eqnarray}
where we used Eq.~(\ref{aaatransfoperator}) of the Appendix and the fact that 
$\{ \hat{L}^\dag_r/\sqrt{2}, \hat{R}_r^\dag/\sqrt{2}\; ; r\}$ is a Kraus set for $\Phi$.
Consequently we can exploit the identity~(\ref{appendix1www}) to write  Eq.~(\ref{imp1111})  
 as 
 \begin{eqnarray}\label{NNNimp1111}
\langle \Psi| \hat{A}^{(s)} |\Psi \rangle = 
  \langle\langle \hat{A}  | (\hat{E}_\Phi)^m | \hat{\rho}_C\rangle \rangle\;.
\end{eqnarray}
Similarly we can proceed for the two-point correlation functions~(\ref{corrimp2}) by defining the
transfer operator of the channel $\Phi\otimes \Phi$ acting on two-triples. Constructing the Liouville space as the tensor product of the 
Liouville spaces of the two triples the latter can be expressed as $\hat{E}_\Phi \otimes \hat{E}_\Phi$, while
Eq.~(\ref{corrimp2})  becomes
 \begin{eqnarray}\label{NNNcorrimp2}
\langle \Psi| \hat{{AB}}^{(s)}_{\Delta_k} |\Psi \rangle = \langle\langle \hat{A} \otimes \hat{B} 
| (\hat{E}_\Phi)^{\bar{m}}  \otimes (\hat{E}_\Phi)^{\bar{m}}  | \hat{\sigma}\rangle\rangle
  \;.
\end{eqnarray}

\section{Thermodynamical limit} \label{sec:term}

In this section we analyze the property of homogenous MERA states in the {\em thermodynamical limit} of $N\rightarrow\infty$.
To approach this problem 
we introduce the family  $\Psi(\chi,\lambda,{\cal C})$ composed by
MERA states  of {\em exponentially} increasing size, i.e.  
\begin{eqnarray}\label{termlim}
\Psi(\chi,\lambda,{\cal C}):=\{ |\Psi_N\rangle : N=2^n \;\; \mbox{for $n\geqslant 3$ integer}\}\;,
\end{eqnarray}
where $|\Psi_N\rangle$ are MERA states with $N$ sites constructed with the same tensors $\chi$, $\lambda$ and ${\cal C}$.
For each one of such family we can then use the result of the previous section to compute the { thermodynamical
limit} of the expectation values of (symmetrized) local observables  as follows
\begin{eqnarray}
A^{(\text{th})}(\Psi) &:=& \lim_{N\rightarrow \infty} \langle \Psi_N | \hat{A}^{(s)} |\Psi_N \rangle \nonumber \\
&=&\lim_{m\rightarrow \infty}   \mbox{{Tr}} [ {\Phi^m}  (\hat{\rho}_C) \;   \hat{A}] \;,
\label{term1}
\end{eqnarray}
with $\hat{A}^{(s)}$ and  $\Phi$ as in 
Sec.~\ref{sec:Qu} and where in the last term we used the fact 
 that for a $N$-sites MERA $m = \log_2 N-2$.
The existence of $A^{(\text{th})}(\Psi)$  depends thus on the convergency of the limit 
$\lim_{m\rightarrow \infty} {\Phi^m}  (\hat{\rho}_C)$, 
with $\Phi$ being the average 
QuMERA channel defined by the tensors $\chi$ and $\lambda$ of the selected family $\Psi$.
This allows us to relate the thermodynamical limit of homogeneous MERA's with the problem of the convergency of repeated applications of a given CPT map and thus ultimately with  its  {\em mixing} (or {\em relaxing}) property~\cite{RAG,GOHM,NJP,TDV}.

It is worth reminding  a mixing channel $\Phi$  is characterized  the property  
\begin{eqnarray}\label{defmix}
\lim_{m\rightarrow \infty} \Phi^m( \hat{B}) = \Phi_f(\hat{B}) := \hat{\rho}_f \; \mbox{Tr} [ \hat{B}] \;,
\end{eqnarray}
with $\Phi_f$ being the CPT map
which  (times $\mbox{Tr} [ \hat{B}]$) transfers every operator $\hat{B}$  into a fix density matrix  $\hat{\rho}_f$
 (the {\em fix point} of $\Phi$).
In the following we will indicate $\Phi_f$ as the {\em final point channel} of $\Phi$. It satisfies the following important property:
\begin{eqnarray}\label{import22}
({\Phi_f} \otimes {\cal I}_Y) (\hat{\Theta}_{XY} ) = \hat{\rho}_f \otimes \hat{\Theta}_Y\;,
\end{eqnarray}
where $\hat{\Theta}_{XY}$ is a joint operator of the system $X$   on which $\Phi_f$  upon  
(i.e. three qudits) and of a generic ancillary system $Y$; ${\cal I}_Y$ is the identity map on $Y$; and finally $\hat{\Theta}_Y := \mbox{Tr}_X[ \hat{\Theta}_{XY}]$. It is a well know
fact that the vast majority of CPT maps acting on given system are mixing (the non-mixing one form a subset of zero-measure).
This clarifies that, a part from some rare pathological case, the limit~(\ref{term1}) is well defined~\cite{NOTAMIX}.
Furthermore it gives us
a simple way of computing such quantity. Indeed owing to Eq.~(\ref{defmix}) we can write
\begin{eqnarray}
A^{(\text{th})}(\Psi) = \mbox{{Tr}} [ \hat{\rho}_f \;   \hat{A}] \;, \label{termsol}
\end{eqnarray}
with $\hat{\rho}_f$ being the fix point of the QuMERA channel $\Phi$. 
The latter can be easily determined by solving the eigenvalue problem
\begin{eqnarray}
\Phi(\hat{\rho})  =  \hat{\rho} \label{eigenva}\;,
\end{eqnarray}
as for mixing maps $\hat{\rho}_f$ is the unique solution of such an equation~\cite{RAG,GOHM,NJP,TDV}.

More generally
we have the following statement
\newline

{\bf Lemma:} {\em The fix point $\hat{\rho}_f$ of the   QuMERA channel $\Phi$ of 
the family $\Psi(\chi,\lambda,{\cal C})$ (when defined) coincides with 
the thermodynamical limit of the average  reduced density matrix of the triple of the associated system, i.e. }
\begin{eqnarray}
\hat{\rho}_f = \lim_{N\rightarrow \infty} \sum_{k=1}^N \hat{\rho}_k/N \;,
\label{opedefrhof}
\end{eqnarray}
{\em with $\hat{\rho}_k$ being the density operator associated with $k$-th triple of $|\Psi\rangle$.}
\newline

{\em Proof:}  exploit the fact that Eq.~(\ref{termsol}) holds for all observables $\hat{A}$ and the fact that
the expectation value of $\hat{A}^{(s)}$ can be expressed as an average over  all triples of the system --- e.g.
see the first line of Eq.~(\ref{proof1}). $\blacksquare$
\newline

{\em Remark:} Since $\Phi$ only depends upon $\chi$ and $\lambda$, the average reduced density matrix 
of the family (as well as the quantities~(\ref{termsol})) 
{\em does not} depend upon the MERA hat tensor ${\cal C}$.
\newline

In a similar way we can also compute the thermodynamic limit of two-point correlation functions~(\ref{ABSIM}) in which
we keep the distance $\Delta_k$ constant.
In particular given
$\hat{{AB}}^{(s)}_{\Delta_k}$ as in Eq.~(\ref{ABSIM}) we define
\begin{eqnarray} \label{term2}
AB^{(\text{th})}_{\Delta_k}(\Psi)& := &\lim_{N\rightarrow \infty} \langle \Psi_N | \hat{AB}_{\Delta_k}^{(s)} |\Psi_N \rangle 
\\&=&\lim_{N\rightarrow \infty}   \mbox{{Tr}} [ ({\Phi^{\bar{m}}\otimes \Phi^{\bar{m}}} ) (\hat{\sigma}) \;   (\hat{A}\otimes \hat{B})] 
\nonumber 
\\&=& \mbox{{Tr}} [ ( {\Phi^{\bar{m}}\otimes \Phi^{\bar{m}}} ) (\hat{\sigma}^{(\text{th})}) \;   (\hat{A}\otimes \hat{B})] \;,
 \nonumber 
\end{eqnarray}
with 
\begin{eqnarray}\label{liminf}
\hat{\sigma}^{(\text{th})} =  \lim_{N\rightarrow\infty} \hat{\sigma} \;,
\end{eqnarray}
being the thermodynamical limit of  $\hat{\sigma}$, the latter being defined as the average reduce density matrix
of $6$ consecutive sites associated with $(\bar{m}+1)$-th MERA layer. Notice that differently from (\ref{term1}) the limit
$N\rightarrow \infty$ does not translate into an infinite sequence of applications of the QuMERA channel $\Phi$. 
This is because the latter  depends only upon the interception among the CS's interception which is fixed by the distance $\Delta_k$. On the contrary varying $N$ implies a variation on $\hat{\sigma}$ which is taken into account by 
Eq.~(\ref{liminf}). By exploiting the scale invariance of homogenous MERA's one can easily verify that 
 $\hat{\sigma}$ coincides with the thermodynamic limit of the average reduced density matrix of $6$ consecutive physical sites
 of the MERA. Therefore  it is possible to show that 
explicit expressions for $\hat{\sigma}^{(\text{th})}$ can be obtained by applying proper (multi-site)
QuMERA channels to $\hat{\rho}_f$. 
In the following section we will not discuss this topic any further,
instead we shall focus on the 
scaling behavior of Eq.~(\ref{term2}) in the limit of large  distances $\Delta_k$.

\section{Scaling behavior of two-point correlations functions}\label{wwwsecscali}

 The scaling  behavior  of the two-points correlations functions for a homogenous MERA
 can be determined  by looking at the spectral  properties of its QuMERA channel $\Phi$~\cite{NOSTRO,nostro1,ALTER}.
This can be done both for finite dimensional MERA's and in the thermodynamical limit thanks to Eqs.~(\ref{corrimp2})  and~(\ref{term2}) which recast  the computation of the correlation functions in term of similar expressions.
Here we will focus on the latter case which is by far the more relevant.

First of all, given $\hat{A}$, $\hat{B}$ generic observables acting on triple of sites, we  introduce the following rescaled quantity
\begin{eqnarray} \label{term3}
\Delta AB^{(\text{th})}_{\Delta_k}(\Psi) &:=& 
AB^{(\text{th})}_{\Delta_k}(\Psi) - A^{(\text{th})}(\Psi)B^{(\text{th})}
(\Psi) \\
&=& \mbox{{Tr}} [ ( {\Phi^{\bar{m}}\otimes \Phi^{\bar{m}}} ) (\hat{\sigma}^{(\text{th})}) \;   (\Delta \hat{A}\otimes \Delta \hat{B})] \;,
 \nonumber 
\end{eqnarray}
where we used the fact that  the 3-site reduced density matrix of $\hat{\sigma}^{(th)}$ is $\hat{\rho}_f$ and where
$\Delta \hat{A}:= \hat{A} - A^{(\text{th})}(\Psi)$, $\Delta \hat{B}:= \hat{B} - B^{(\text{th})}(\Psi)$.
The idea is to invoke once more the mixing properties of~$\Phi$ which guarantee that
this operator has a unique unitary eigenvalue~\cite{RAG,GOHM,NJP,TDV}. Exploiting  then  the spectral decomposition 
of  $\Phi$ and keeping the highest order contributions this can now be written as~\cite{NOSTRO},
\begin{eqnarray} \label{term4}
\Delta AB^{(\text{th})}_{\Delta_k}(\Psi)\Big|_{\bar{m}\gg1}  \simeq
 c \; |\eta \eta'|^{\bar{m}} \simeq c \;  \Delta_k^{\log_2 |\eta \eta'|}  \;, 
\end{eqnarray}
with $\eta$, $\eta'\neq 1$ being the  eigenvalues of $\Phi$ of largest modulus which contribute non trivially in the expansion,
and where in the last term we used the fact that  $\bar{m}$ scales logarithmically with the distance
  $\Delta_k$ as in Eq.~(\ref{Delta}). In this expression $c$ is a term which  scales at most polynomially on $m$, i.e. $c\simeq {\cal O}(\text{Poly}(m)) \simeq {\cal O}(\text{Poly}(\log_2 \Delta_k))$  --- see below. Equation~(\ref{term4}) shows a polynomial decay of the two-point correlation function of the system which is
typical of critical system~\cite{ZINN}.
 Its derivation  resembles a similar calculation performed
in Ref.~\cite{WOVC} for MPS's. It can be obtained by 
expressing $\Delta AB^{(\text{th})}_{\Delta_k}(\Psi)$ in
the Liouville representation which as seen in Sec.~\ref{sec:to} gives 
\begin{eqnarray} \label{term3}
\Delta AB^{(\text{th})}_{\Delta_k}(\Psi) \label{fea}
= \langle\langle {\Delta \hat{A}} \otimes {\Delta \hat{B}} | 
 (\hat{E}_{\Phi})^{\bar{m}} \otimes (\hat{E}_{\Phi})^{\bar{m}}  | \hat{\sigma}^{(\text{th})} \rangle\rangle,
\end{eqnarray}
where  
 $|\hat{\sigma}^{(th)}\rangle\rangle$ is the vector of 
  $\left({\cal H}_d^{\otimes 3} \otimes {\cal H}_d^{\otimes 3} \right)^{\otimes 2}$ which 
  represent the state $\hat{\sigma}^{(th)}$.
 This can now be simplified by means of the identity Eq.~(\ref{cuc1}) and 
  observing that for each vector $|\hat{\sigma} \rangle\rangle\in
 \left({\cal H}_d^{\otimes 3} \otimes {\cal H}_d^{\otimes 3} \right)^{\otimes 2}$ the following identities
 applies
 \begin{eqnarray}
 \langle\langle {\Delta \hat{A}} \otimes {\Delta \hat{B}} | 
(\hat{E}_{f} \otimes\hat{\openone}) | \hat{\sigma} \rangle\rangle &=&
\langle\langle {\Delta \hat{A}} \otimes {\Delta \hat{B}} | 
({\Phi} \otimes {\cal I} ) (\hat{\sigma}) \rangle\rangle \nonumber \\
= \langle\langle {\Delta \hat{A}} \otimes {\Delta \hat{B}} | 
\hat{\rho}_f \otimes \hat{\rho} \rangle\rangle &=& \langle\langle {\Delta \hat{A}}  | 
\hat{\rho}_f  \rangle\rangle \langle \langle  {\Delta \hat{B}}| \hat{\rho} \rangle\rangle \nonumber \\
= \mbox{Tr} [ \Delta \hat{A} \; \hat{\rho}_f ] \; \mbox{Tr}  [ \Delta \hat{B} \; \hat{\rho} ] &=&0\;,
\label{uno}
 \end{eqnarray}
 (here $\hat{\rho}$ is the reduced density matrix of  
 $\hat{\sigma}$ and  we used the property Eq.~(\ref{defmix}) of $\Phi$, 
 and the fact that $\mbox{Tr} [ \Delta \hat{A} \; \hat{\rho}_f ] =0$).
 Similarly one has
  \begin{eqnarray}
 \langle\langle {\Delta \hat{A}} \otimes {\Delta \hat{B}} | 
(\hat{\openone} \otimes \hat{E}_{f}) | \hat{\sigma} \rangle\rangle =0 \;.
\label{due}
 \end{eqnarray}
Exploiting these identities we can now write 
 \begin{eqnarray}\label{tre}
\Delta AB^{(\text{th})}_{\Delta_k}(\Psi) &=&  \langle\langle {\Delta \hat{A}} \otimes {\Delta \hat{B}} | 
 ( \Delta\hat{E}_{\bar{m}} \otimes  \Delta\hat{E}_{\bar{m}}) | \hat{\sigma}^{(\text{th})} \rangle\rangle
 \nonumber \\
 &=& \sum_{j,j' \neq 0}  |\eta_{j}
\eta_{j'} |^{\bar{m}} \; C_{j j'}^{(\bar m)} \;,\end{eqnarray}
where the  $C_{jj'}^{(\bar m)}$ being trigonometric, polynomial functions of $\bar{m}$ defined by
\begin{eqnarray}
C_{jj'}^{(\bar m)} := 
\; \langle\langle\Delta \hat{A} \otimes \Delta \hat{B}  | \hat{e}_j(\bar{m}) \otimes \hat{e}_{j'} (\bar{m}) 
 | \hat{\sigma}^{(th)} \rangle\rangle\;,
\end{eqnarray}
with $\hat{e}_{j,j'}(\bar{m})$ as in Eq.~(\ref{wwwdefe}).
 Equation~(\ref{term4}) finally follows 
by  taking the couple $j,j'$ which has the largest value of $|\eta_{j}
\eta_{j'} |$ and for which $C_{jj'}^{(\bar m)}\neq 0$. 

\subsection{Self-adjoint transfer super-operator} \label{self}
Of special interest is the case of MERA's which have a self-adjont  $\Phi$ 
QuMERA channel~(\ref{QUMERA1}) 
(i.e. $\Phi = \Phi_H$ with $\Phi_H$ representing $\Phi$ in Heisenberg picture).
An example of such MERA's has been recently studied in Ref.~\cite{ALTER} in the
calculation of the ground state properties of Ising and Pootz model. 
In this case we can write
\begin{eqnarray}
\Phi(\hat{A}) = \sum_{j} \eta_j \; \mbox{Tr} [\hat{\Theta}_j^\dag \hat{A}] \; \hat{\Theta}_j
\end{eqnarray}
with $\eta_j$ being the (real) eigenvalues of $\Phi$ and $\hat{\Theta}_j$ being
the corresponding eigen-operator properly orthonormalized with respect to the Hilbert-Smidth
scalar product (in particular if $\Phi$ is mixing then $\eta_0=1$ is non-degenerate and $\hat{\Theta}_{j=0} = \hat{\rho}_f$). Therefore Eq.~(\ref{term3}) yields
\begin{eqnarray} \label{term3333}
\Delta AB^{(\text{th})}_{\Delta_k}(\Psi) 
&=& \sum_{j,j' \neq 0} \eta_j^{\bar{m}}  \eta_{j'}^{\bar{m}}  \;\; 
\mbox{{Tr}} [ (\hat{\Theta}_j^\dag\otimes \hat{\Theta}_{j'}^\dag)   (\hat{\sigma}^{(\text{th})}) ] \nonumber \\
&\times &\; \mbox{{Tr}} [ (\hat{\Theta}_j\otimes \hat{\Theta}_{j'})   (\Delta \hat{A}\otimes \Delta \hat{B})]\;,
\end{eqnarray}
where the properties (\ref{uno}) and (\ref{due}) has been used to remove the contributions in $j=0$ or 
$j'=0$ from the sum.
 The above expression  coincides with Eq.~(\ref{tre}) by identifying $C_{jj'}^{(\bar m)}$ with 
the coefficients $\mbox{{Tr}} [ (\hat{\Theta}_j^\dag\otimes \hat{\Theta}_{j'}^\dag)   (\hat{\sigma}^{(\text{th})}) ] \mbox{{Tr}} [ (\hat{\Theta}_j\otimes \hat{\Theta}_{j'})   (\Delta \hat{A}\otimes \Delta \hat{B})]$ times a phase factor. A direct proof of this can easily be obtained by observing that in this case 
the Liouville representation of $\Phi$ is provided by the Hermitian operator
\begin{eqnarray}
\hat{E}_\Phi  = \sum_{j} \eta_j \; |\hat{\Theta}_j \rangle\rangle \langle\langle \hat{\Theta}_j| \;,
\end{eqnarray}
with $|\hat{\Theta}_j\rangle\rangle$ being orthonormal.
Under this condition the power-law scaling~(\ref{term4}) becomes exact as the coefficient 
$c$ is now independent from $\bar{m}$.
 
\section{Conclusions} \label{sec:conc}
In this paper we have presented an extensive analysis of the properties of 
homogeneous MERA states
based on the quantum channels approach introduced in Ref.~\cite{NOSTRO}.
In particular we have proved some  Theorems which allows us to characterized the
(local) average properties of such states in terms of the spectrum of a single QuMERA channel $\Phi$:
Theorem 1 establishes that the expectation value of any (average) local observable can be evaluated on the fix point (eigenvector associated with the maximum eigenvalue) of $\Phi$;
Theorem 2 instead gives an explicit expression for the two-points correlation functions of the system. Both Theorems holds also in the thermodynamic limit of MERA states associated
with an infinite number of sites. In particular Theorem 2 allows one to identify the (power law)
scaling behavior of the MERA.  For the sake of simplicity
 the analysis has been performed assuming a specific MERA decomposition but it 
 can be trivially generalized to any possible variation of the latter.

\acknowledgments 
This work was in part founded by the Quantum Information
research program of Centro di Ricerca Matematica Ennio De Giorgi
of Scuola Normale Superiore. 

\appendix

\section{Liouville space representation}\label{Liou}

As anticipated in Sec.~\ref{sec:to} 
the Liouvillle representation  is constructed by fixing an orthonormal  basis $\{ |i\rangle; i\}$ 
on the Hilbert space of interest ${\cal H}$ 
 (e.g. the three qubits space ${\cal H}_d^{\otimes 3}$) 
and by defining the following {\em linear} mapping 
from the space ${\cal B}({\cal H})$ of the linear operators of  ${\cal H}$ to 
${\cal H}\otimes {\cal H}$, 
\begin{eqnarray}
\hat{A} = \sum_{ii'} \langle i| \hat{A}|i'\rangle \; |i\rangle\langle i'|   \rightarrow 
| \hat{A} \rangle\rangle := \sum_{ii'}  \langle i| \hat{A}|i'\rangle |i\rangle\otimes | i' \rangle 
\label{dueapp}.
\end{eqnarray}
Simple but useful properties of the mapping~(\ref{dueapp}) are the following rules,
\begin{eqnarray}
| \hat{A} {\hat B} {\hat C}\rangle \rangle &=& (\hat{A} \otimes \hat{C}^T ) | \hat{ B }\rangle \rangle \label{faf}\;, \\
\mbox{Tr} [ \hat{A}^\dag \hat{B}] &=& \langle\langle \hat{A} | \hat{B} \rangle\rangle \;, 
\end{eqnarray}
which hold for all operator $\hat{A}, \hat{B} , \hat{C} \in {\cal B}({\cal H})$. In these expressions $\hat{C}^T$ stands for transposition {\em with respect to the selected basis} $\{ |i\rangle ; i\}$, while $\hat{A}^\dag$ is the  adjont of $\hat{A}$ -- notice that 
one has $\langle \langle \hat{A} | = [| \hat{A}\rangle\rangle ]^\dag = \sum_{ii'}  \langle i| \hat{A}|i'\rangle^* \langle i | \otimes \langle i' |$ with $\langle i| \hat{A}|i'\rangle^*$ being
the complex conjugate of $\langle i| \hat{A}|i'\rangle$.
It is also worth noticing that, according to the above expressions, the vector $|\hat{\openone} \rangle\rangle = \sum_{i} |i\rangle\otimes |i\rangle$ 
satisfies the following identities
\begin{eqnarray}\label{traccia}
\mbox{Tr} [ \hat{B}] &=& \langle\langle \hat{\openone} | \hat{B} \rangle\rangle \;,
\\
|\hat{B} \rangle\rangle &=& (\hat{B} \otimes \hat{\openone} ) | \hat{\openone}\rangle \rangle = 
 (\hat{\openone} \otimes\hat{B}^T) | \hat{\openone}\rangle \rangle\;.
\end{eqnarray}
In this language a CPT map  ${\Phi}$  operating on ${\cal B}({\cal H})$ 
is described by an operator acting on
${\cal H} \otimes {\cal H}$ defined by
\begin{eqnarray}\label{aaatransfoperator}
\hat{E}_\Phi := \sum_s \hat{ M}_s \otimes \hat{M}_s^*\;,
\end{eqnarray}
where $\hat{M}_s$  are a set of Kraus operators of $\Phi$.
Equation~(\ref{aaatransfoperator})
 is a consequence of (\ref{faf}) by noticing that
for all $\hat{B} \in {\cal B}({\cal H})$ one has
\begin{eqnarray}
| \Phi(\hat{B}) \rangle\rangle 
&=& \sum_s |\hat{M}_s \hat{B} \hat{M}^\dag_s \rangle\rangle = \label{NNWW}
\sum_s  \hat{M}_s\otimes \hat{M}^*_s |  \hat{B}  \rangle\rangle \nonumber \\
&=& \hat{E}_\Phi  |\hat{B}\rangle\rangle \;.
\end{eqnarray}
This expression shows  that $\Phi$ and its associated matrix $\hat{E}_\Phi$ have the same spectrum (i.e. the same eigenvalues). It also allows us to express 
the expectation values on evolved operators as matrix elements of $\hat{E}_\Phi$  
as indicated by the following expression
\begin{eqnarray}\label{appendix1www}
\mbox{Tr} [ \hat{A}^\dag \Phi(\hat{B})] &=& \langle\langle \hat{A} | 
\hat{E}_\Phi | \hat{B} \rangle\rangle \;.
\end{eqnarray}
Finally it allows one to compute the successive application of a CPT map as follows:
\begin{eqnarray}\label{multi}
| \Phi^m(\hat{B}) \rangle\rangle  = (\hat{E}_\Phi)^m  |\hat{B}\rangle\rangle \;.
\end{eqnarray}
An interesting problem  is to determine the limit for $m\rightarrow \infty$ of $(\hat{E}_\Phi)^m$ when $\Phi$ is mixing (as in the case of the QuMERA channel case).
From Eq.~(\ref{defmix}) we know that this must be the transfer matrix $\hat{E}_{\Phi_f}$ of the channel  $\Phi_f$ which maps every operator into the fix point $\hat{\rho}_f$. According to the above definitions this implies
\begin{eqnarray}
\hat{E}_{\Phi_f} | \hat{A}\rangle\rangle = \mbox{Tr} [ \hat{A}]  \; |\hat{\rho}_f\rangle\rangle = 
\langle\langle \hat{\openone} | \hat{A}\rangle \rangle \; |\hat{\rho}_f\rangle\rangle\;.
\end{eqnarray}
Since this must be true for all vectors $|\hat{A}\rangle\rangle$ we can conclude
\begin{eqnarray}\label{import0}
\lim_{m\rightarrow \infty} (\hat{E}_\Phi)^m = \hat{E}_{\Phi_f} = | \hat{\rho}_f \rangle\rangle \langle \langle \hat{\openone} |\;,
\end{eqnarray}
which is  consistent with Eq.~(\ref{termsol}).

\subsection{Some facts about the spectrum of $\Phi$ and $\hat{E}_\Phi$}
Consider the set ${\cal S}(\Phi) := \{ \eta_j ; j\}$ of the eigenvalues of $\Phi$. 
They are defined by eigenvector equations of the following form
\begin{eqnarray}\label{lab}
\Phi(\hat{\Theta}_j)  = \eta_j \hat{\Theta}_j \quad \Longleftrightarrow  \quad 
\hat{E}_\Phi |\hat{\Theta}_j \rangle\rangle   = \eta_j |\hat{\Theta}_j \rangle\rangle\;,
\end{eqnarray}
where $| \hat{\Theta}_j\rangle\rangle$ are the vectors associated with  the  eigenvectors 
$\hat{\Theta}_j$ of $\Phi$ (for each $\eta_j$ there can be 
more than one  $|\hat{\Theta}_j\rangle\rangle$).
Since the matrix  $\hat{E}_\Phi$ is generally not Hermintian the $k_j$ will  be not real. However since $\Phi$ is CPT one can show that
$\eta_j$ belongs to unit circle~\cite{BENGZY}
(i.e. $| \eta_j | \leqslant 1$),
and that $\eta_0:= 1$ is always an element of the spectrum, i.e.  $1 \in {\cal S} (\Phi)$.
Furthermore one has that since $\Phi$ is trace preserving then the eigenvectors associated to eigenvalues $\neq 1$
are traceless operators, i.e. $\mbox{Tr} [ \hat{\Theta}_j ]=0$ (in particular they cannot be density matrices).
Finally one can verify that if  $\eta_j\in {\cal S}(\Phi)$ than also its c.c. is an eigenvalue of $\Phi$, i.e. 
 $\eta_j^* \in {\cal S}(\Phi)$ (too see this just take the adjoint of the left hand side equation of~(\ref{lab})
 and use the fact that $\Phi(\hat{\Theta})^\dag = \Phi(\hat{\Theta}^\dag)$).
 More generally one can verify that the couple  $\eta_j$ and $\eta_j^*$ have the same Jordan structure,
 i.e. their corresponding Jordan blocks (see below) will have the same dimensions.

As already mentioned the spectral properties of $\Phi$ determine uniquely the mixing property of the map~\cite{RAG,GOHM,NJP,TDV}.
In particular, it is known that the map is {\em ergodic} (i.e.  it has a unique fix point that satisfies Eq.~(\ref{termsol}))
 if and only if  $\eta_0$ is non-degenerate (i.e. it has a unique eigenvector).
Furthermore, it is known that 
$\Phi$ is {\em mixing} if and only if $\eta_0=1$ is the only  eigenvalue with unitary modulus and it 
is (non-degenerate),
i.e. $|\eta_j |=1$ iff $\eta_j = 1$.
We finally remind that ergodic maps are not necessarily mixing even though any mixing channel is necessarily ergodic, and that
mixing channels are dense in the set of the CPT maps.

\subsection{Jordan block decomposition of $\hat{E}_\Phi$}

We have seen that in general $\hat{E}_\Phi$ is not Hermitian: as a matter of fact, typically it will not be even orthogonal i.e. diagonalizable -- see however the discussion of Sec.~\ref{self} and Ref.~\cite{ALTER}.
 We can however still put it in Jordan form by similarity transformation, i.e. 
\begin{eqnarray} \label{decJ0}
\hat{E}_\Phi = \hat{T} \hat{J} \hat{T}^{-1}\;,
\end{eqnarray}
where $\hat{T}$ is an invertible operator and where $\hat{J}$ is the Jordan form associated with $\hat{E}_\Phi$. A part from a trivial permutation of the blocks, the operator $\hat{J}$ is uniquely
determined as
$\hat{J} := \oplus_j \hat{J}_{d_j}(\eta_j)$ 
with $\hat{J}_{d_j}(\eta_j)$ being the Jordan block of dimension $d_j$ associated with the eigenvalue $\eta_j$ of $\Phi$.
It is worth reminding that each eigenvalue  $\eta_j$ can have more than a single block:  the total number of such blocks
corresponds to the so called {\em geometric multiplicity} of $\eta_j$, i.e. to the number of linearly independent eigenvectors of $\eta_j$ (i.e. to the dimension of the associated eigenspace). Finally the sum of the dimension $d_j$ of all the blocks associated with a given
eigenvalue $\eta_j$ is to the so called {\em algebraic multiplicity} of $\eta_j$, i.e.  the number of zeros of the characteristic polynomial associated with the
solution $x= \eta_j$~\cite{HORN}. For  diagonalizable matrices one has $d_j =1$ and the geometric multiplicity coincides with the algebraic one. 
We remind  also that the  operator $\hat{J}_{d_j}(\eta_j)$ can be written as the following matrix
\begin{eqnarray}\label{decJ}
\hat{J}_{d_j}(\eta_j) = \eta_j \hat{\openone}_{d_j} + \hat{N}_{d_j}\;,
\end{eqnarray}
with $\hat{\openone}_{d_j}$ being the $d_j \times d_j$ identity matrix and where $\hat{N}_{d_j}$ is
nilpotent matrix which satisfy the condition $(\hat{N}_{d_j})^p = 0$ for all $p\geqslant d_j$
(specifically it is
either  
a  $d_j \times d_j$ 
matrix of $1$'s above the diagonal or is the null matrix).

  It is interesting to observe that the matrix $\hat{E}_\Phi$ satisfies the following condition
 \begin{eqnarray}
 \hat{E}_\Phi^* = \sum_s \hat{M}_s^* \otimes \hat{M}_s = \hat{S} (\hat{E}_\Phi) \hat{S}^\dag \label{sss}
 \end{eqnarray}
 where $\hat{S}$ is the swap operator which sends any operator of the form  $\hat{A}\otimes \hat{B}$ 
 into $\hat{B}\otimes \hat{A}$. This is unitary and Hermitian (i.e. $\hat{S}= \hat{S}^\dag = \hat{S}^{-1}$). Therefore 
 $\hat{E}_\Phi^*$ and $\hat{E}_\Phi$ are
 connected through a  similarity transformation (i.e.  they are mapped into each other by an invertible operator): 
 consequently $\hat{E}_\Phi$ and $\hat{E}_\Phi^*$ admit the same Jordan form decomposition.
 This is an important fact that tell us that given an complex conjugate couple  $\eta_j$ and $\eta_j^*$ of eigenvalues of
 $\hat{E}_\Phi$, their associated Jordan block will have the  same structure (i.e.  dimension and possible degeneracies).
From this point of view hence $\hat{E}_\Phi$ share some properties  the real  matrix. In particular one can decompose $\hat{E}_\Phi$ in the so called
{\em real Jordan form} where the (complex) blocks associated with each couple of complex conjugate eigenvalues can be grouped together to form real "super-blocks"~\cite{HORN}.
Finally  it is worth remembering that $\hat{E}_\Phi$ is connected through similar transformation also with its transpose $\hat{E}_\Phi^T$ (this is a general property
of all complex matrices~\cite{HORN}), i.e. $\hat{E}_\Phi = \hat{R} \; \hat{E}_\Phi^T \; \hat{R}^{-1}$ with $\hat{R}$ invertible but not necessarily unitary. Furthermore since 
$\hat{E}_\Phi^T$ is connected through $\hat{E}_\Phi^\dag$ by swaps operation (the derivation is as in Eq.~(\ref{sss})) it follows that these all these 
matrices are  similarly equivalent i.e. 
\begin{eqnarray}
\hat{E}_\Phi \sim \hat{E}_\Phi^* \sim \hat{E}_\Phi^T \sim \hat{E}_\Phi^\dag \label{similarity}\;,
\end{eqnarray}
(where ``$\sim$" stands for the similarity equivalence), and will have the same spectra~\cite{NOTAIMPO}.

 The above expressions are extremely useful when computing successive application of $\Phi$ -- see Eq.~(\ref{multi}). Indeed 
 from Eqs.~(\ref{decJ0}) and (\ref{decJ}) one gets
\begin{eqnarray}\label{cuc}
( \hat{E}_\Phi)^m = \hat{T}  \hat{J}^m  \hat{T}^{-1} =  \hat{T} \Big(\oplus_j \; 
[ \hat{J}_{d_j}(\eta_j) ]^m 
\Big) \hat{T}^{-1} \;,
\end{eqnarray}
with $[ \hat{J}_{d_j}(\eta_j)]^m = \eta_j^m \; \hat{Q}^{(m)}(\eta_j)$ and 
 \begin{eqnarray}
\hat{Q}^{(m)}(\eta_j) 
:= \sum_{q=0}^{d_j -1} \left( \begin{array}{c} m \\ q\end{array} \right) \; \eta_j^{-q} \; (\hat{N}_{d_j})^q  ,
\label{quirc}
\end{eqnarray}
 being 
  bounded operators which are polynomial in $m$.
 This implies that for $|\eta_j |<1$ one has 
  \begin{eqnarray}
\lim_{m\rightarrow\infty}  
[\hat{J}_{d_j}(\eta_j)]^m 
=
 \hat{\O}_{d_j}\;,
\end{eqnarray}
with $\hat{\O}_{d_j}$ being the $d_j \times d_j$ null matrix.
Therefore for $\Phi$ mixing we can write
\begin{eqnarray} \label{import1}
\lim_{m \rightarrow\infty} [\hat{E}_\Phi]^m
=
\hat{T} \Big(  \hat{J}_{1}(\eta_0)  \oplus_{j \neq 0}  \hat{\O}_{d_j} \Big) \hat{T}^{-1} = \hat{E}_{\Phi_f},
\end{eqnarray}
where  we use Eq.~(\ref{import0}) and the fact 
  that for a mixing channel  $\eta_0=1$ is not degenerate and thus its corresponding 
Jordan block $\hat{J}_{1}(\eta_0)$ is the $1\times 1$ matrix formed by the single element $1$.
 Thus Eq.~(\ref{cuc}) can now be written as follows
\begin{eqnarray}\label{cuc1}
(\hat{E}_\Phi)^m 
= \hat{E}_{\Phi_f} +  \Delta_m \hat{E}_{\Phi}
\;,
\end{eqnarray}
where $\Delta_m \hat{E}_{\Phi}$ is a contribution that nullifies for $m\rightarrow \infty$. It can be expressed as
\begin{eqnarray} \label{lld}
 \Delta_m \hat{E}_{\Phi} : = \sum_{j\neq 0} |\eta_j|^{m} \hat{e}_j(m)\;,
 \end{eqnarray}
 with the matrix $\hat{e}_j(m)$ being a trigonometric, polynomial function defined by 
\begin{eqnarray}\label{wwwdefe}
\hat{e}_j(m) := \hat{T} [ \oplus_{j\neq 0}   e^{i m \arg[\eta_j]} \;  \hat{Q}^{(m)}(\eta_j)] \hat{T}^{-1}\;.
  \end{eqnarray}
 It is finally worth mentioning that for Hermitian $\hat{E}_\Phi$ 
the whole analysis simplify. In this case in fact $\hat{T}$ is a unitary transformation, the $\eta_j$'s are real,  while 
$\hat{Q}^{(m)}(\eta_j) = \hat{\openone}_{d_j}$. Under this condition $\hat{e}_j(m)$ become
independent from $m$ and coincides with projector on $j$-eigenspace  of $\hat{E}_{\Phi}$.


\begin{thebibliography}{99}
\bibitem{NOTAnew1} This is not a serious  limitation as one can arbitrarily  
increase the local dimensions $d$ of the physical indexes of the system.
\bibitem{NOTACONTRACT} 
Given ${\cal A}_{1}$ and ${\cal A}_2$  tensors of 
type-$\mbox{\tiny{$\left(\begin{array}{c} M_{1} \\ N_{1} \end{array}\right)$}}$
and type-$\mbox{\tiny{$\left(\begin{array}{c} M_{2} \\ N_{2} \end{array}\right)$}}$, respectively,  
${\cal A}_1 \cdot {\cal A}_2$ is the tensors obtained  by contracting the upper indexes of ${\cal A}_1$ with the lower
 indexes of ${\cal A}_2$. When not specified otherwise we assume that such contraction is performed
 by coupling every upper index of ${\cal A}_1$ with the corresponding lower one of ${\cal A}_2$.
 Thus for $M_1=N_2$ this implies  ${\cal A}_1 \cdot {\cal A}_2 :=   [{\cal A}_1]^{\circ,\cdots, \bullet}_{\ell_1,\cdots,\ell_{N_1}} \; [{\cal A}_2]_{\circ,\cdots, \bullet}^{u'_1, \cdots, u'_{M_2}}$. Similarly if the upper indexes of ${\cal A}_1$ exceed  the lower indexes of ${\cal A}_2$ or vice-versa the saturation
 will involve only the available couples, i.e. 
 ${\cal A}_1 \cdot {\cal A}_2 :=   [{\cal A}_1]^{\circ,\cdots, \bullet}_{\ell_1,\cdots,\ell_{N_1}} \; [{\cal A}_2]_{\circ,\cdots, \bullet, \ell'_{M_1+1}, \cdots, \ell'_{N_2}}^{u'_1, \cdots, u'_{M_2}}$ for $M_1<N_2$, and 
  ${\cal A}_1 \cdot {\cal A}_2 :=   [{\cal A}_1]^{\circ,\cdots, \bullet, u_{N_2}, \cdots, u_{M_1}}_{\ell_1,\cdots,\ell_{N_1}} \; [{\cal A}_2]_{\circ,\cdots, \bullet}^{u'_1, \cdots, u'_{M_2}}$ for $M_1>N_2$. 
 When needed we will use brackets to specify how to treat concatenating  sequences of ``$\cdot$"-product.
\bibitem{NOTA3} The special role played by the compounds of 3 consecutive sites, is related with the 
topology of the MERA network we are analyzing here. In alternative schemes~\cite{ALTER}  the triples are replaced by
other compounds.
\bibitem{NOTATENSO} 
This is the type-$\mbox{\tiny{$\left(\begin{array}{c} 3 \\3 \end{array}\right)$}}$ tensor of element
elements $[{A}_k]_{\ell_1,\ell_2,\ell_{3}}^{u_1,u_2, u_3}:= \langle  \xi_{u_1},  \xi_{u_2} , \xi_{u_3}  | \hat{A}_k | 
\xi_{\ell_1}, \xi_{\ell_2},\xi_{\ell_3} \rangle$.
\bibitem{NOTAIMPO} Exploiting the  property that similar matrices share the same spectra and Eq.~(\ref{similarity}) it is easy to verify that the maps $\Phi^{(R)}$ and 
$\Phi^{(L)}$ 
have the same spectra. This implies in particular that if one of the two is mixing also the other will be mixing -- see Sec.~\ref{sec:term}.
\bibitem{NOTAPERO} This does not prevent  the possibility that in the thermodynamic limit 
$N\rightarrow \infty$ homogenous  MERA's could  become asymptotically translational invariant.
\bibitem{NOTAELEMENTO}Without loss of generality in the following we shall assume $\{ |i\rangle; i\}$ to be the computational basis of the 3 three qudits. With this choice the matrix $\langle i| \hat{A}|i'\rangle$ coincides with the
tensor~\cite{NOTATENSO} of the operator.
\bibitem{NOTAMIX} It is worth noticing that
the mixing property of $\Phi$ is a {\em sufficient} condition for the existence of the limit 
(\ref{term1}), i.e. the latter could still be well defined also for families $\Psi(\chi,\lambda,{\cal C})$ which
are  associated with a NON mixing QuMERA channel (the definition of $A^{(\text{th})}(\chi,\lambda,{\cal C})$ only requires
that $\Phi^m$ converges when applied to a specific state $\hat{\rho}_C$).
\end{thebibliography}
\end{document}